# SecureIT using Firebase, Google map and Node.Js

Shanta Khatun[1][0000-0002-9810-5761], Fahim Hossain Saiki[2][0000-0002-1416-460X] and Milon Biswas[3][0000-0002-2825-8308]

[1] Bangladesh University of Business and Technology, Dhaka, Bangladesh
shanta.cse.bubt@gmail.com
[2] Shahjalal University of Science and Technology, Sylhet, Bangladesh
fahim91@student.sust.edu
[3] Bangladesh University of Business and Technology, Dhaka, Bangladesh
milonbiswas@bubt.edu.bd

**Abstract.** This paper is about describing the features of a software that was developed for its user's safety which we called "SecureIT" is a android based software using Android SDK along with Firebase and Google map SDK along with Node.Js. The aim of developing this project was to make sure and taking its users safety to a next level. Actually now a days some crime incidents like rapes, fire accidents and snatchings are very common and we believe many of those can be prevented if victim got support at the right time. According to the well known daily news paper of Bangladesh "The Daily Star" there were about 1413 women was rapped where 76 women were dead in 2019. On the same pa-per it also said that in 2018 and 2017 the number of rapes were 732 and 818 where we can easily get that the number increases to almost double in 2019. Where we get a point that if those girls get support or get people known about their location at the right time they might get rid of the situation and the number of rapes could be reduced a lot because we all know that now a days using mo-bile smartphone is too easy for people. Although leading Chinese mobile phone company Xiaomi introduced a new feature "Emergency SOS service" in their mobile phones at the end of 2018 with a MIUI 10 update but this feature is only limited to the Xiaomi phones and it was well advertised as well. This paper will briefly describe all the features of our software and its usages.

**Keywords:** Android, SDK, Firebase, Google map, Xiaomi, Node.JS, MIUI.

## 1    Introduction

Now a days people safety is a very big concern in this society. If we open newspaper daily, we must find some news on crimes like rapes, fire incidents etc. In our country the most identical crimes happening again and again are rapes and fire accidents which are taking many people to death and destroy their family's future too. But it must be noticed if there will some way to let the public administrations like police, fire service know about the possibilities of getting affected by any of those incidents or making family members or known people alert about their location and situation then many of those crime can be prevented and also many of lives and their dreams can be made safe and secure and for making this possible this product was developed.

The leading Chinese mobile telecom company Xiaomi has introduced their new feature called "Emergency SOS" in late 2018 with the stable update of MIUI 10 . But unfortunately this feature was not advertised well and most of the Xiaomi users does not even know about the feature and also this feature is not too user friendly as a user has to press the power button rapidly for the 5 times to send a SOS message to user's fixed contacts and this feature is limited to that but in "SecureIT" a user also has the access to view the nearest hospitals, fire services and police stations by the help of Google map SDK .

Our product is only not limited to that it also ensure its best services to protect its users as they have the option to chat with people when a user finds no help from his/her trusted contacts which is developed with the help of "Firebase Realtime Data-base" and also the data that to be taken from a user at the time of registration are secured with one of the best authentication system out there which is "Firebase authentication system" .

This project also uses Firebase messaging service , Firebase function and Node.js which automate the delivery of the notification about getting text message with build in online chat system using Firebase real-time database in this software which makes the software more user friendly to its user .This was also the magic of Firebase SDK where the software will collect Firebase token at the starting as it will be used for targeting a device to send the notification.

The main inspiration behind this development was if those crimes were being analyzed well we will find most of those victims are middle aged and modern enough to operate an android based smartphones and analysis of those crime scene also says that most of them used a smartphone So, our aim for this project was to make a user's smartphone as his/her best weapon to protect himself/herself from getting affected by any of those crimes..

With a reason of the making this software available to everyone the minimum SDK version for running this software on an device is set to 16 which is Android 4.1 (Jelly Bean) and also "Sign up with Facebook and Gmail" feature using Firebase SDK is implemented in this software so that maximum number of android device can run this software with the most user-friendly way.

## 2        Technologies and Methodologies used

In this section we will dissect some of the major technologies that is used for developing our "SecureIT". Normally we used Android SDK and also every support provided by Android studio.

### 2.1    Android SDK

It is a software development kit which allows to develop software for android platforms. It includes sample project with tools, source code, emulators and all the required support libraries for the development of an android based application. Most of these libraries come from google, maven and jcenter which are basically defined in project level gradle file in Android studio. Most of the developments with Android SDK is generally written in Kotlin, Java and C++ although languages like Go,    JavaScript, C, C++ or assembly also used for developing with Android SDK partially.

## 2.2    XML

XML which is Extensible Markup Language is actually defined as the collection of rules for encoding some records in a fixed format which is both human-readable and machine-readable. XML is specified mostly as World Wide Web Consortium's XML 1.0 specification which is introduced in 1998. Although there are many related open standard specifications are available out there for Extensible Markup Language.

The aim of introducing XML or Extensible Markup Language was to emphasis simplicity, generality and handy across the internet. It is a textual data format with stable support via Unicode for different types of languages across the internet. Although the main focus of the design of the XML is document but the most of the developers use this language for the representation of arbitrary data structures for instance which are used in web services.

## 2.3    Android Studio

For the development of an android application the official IDE from Google and JetBrains is a IntelliJ IDEA based software which is Android studio. Actually, Android studio is written in Java, Kotlin and C++ and the first stable version of Android studio was released in June, 2014. Android Studio is the best for android software development because of its services. Basically, Android studio offers :

1. Gradle based building system
2. Various Apk file generation
3. Proguard rules.
4. App signature capabilities
5. Keystore generation during release
6. Rich editor for different types of layout including drag and drop, etc.

## 2.4    Java

JAVA is a programming language which is developed for the environment of internet. The architecture of Java is mostly like C++ Language. It is said that Java has the "look and feel" of the C++ but it is too easy than C++ and also its emphasis an object-oriented programming model. It is one of those languages that is used for    developing a whole application in a single language and can be executed on a single machine.

## 3 Firebase SDK

For development of this software we needed some features to make the software most user friendly and for this we used Firebase SDK and these are those parts of Firebase we used for SecureIT bellow.

### 3.1 Firebase Authentication system

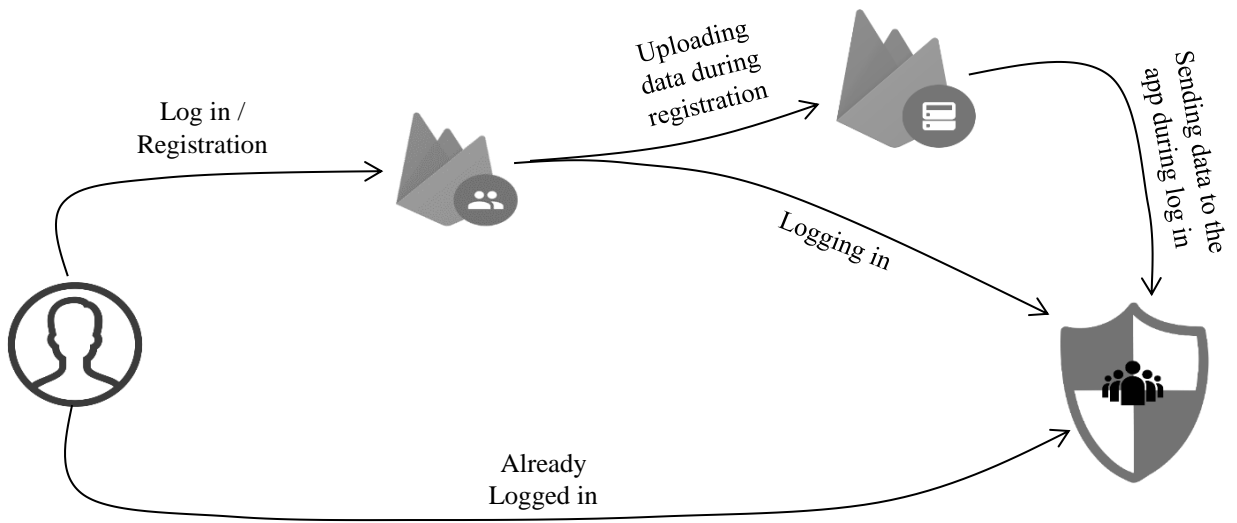

**Fig. 1.** Firebase authentication system

In Fig. 1 the main entrance way of our software is shown and which is restricted by Firebase Authentication system and also its functionality where users stores the media like display picture in firebase storage and other data in in Firebase real-time database during registration and after that users must verify their email to use the software and after verifying their email they can enter in "SecureIT" by logging in . So, it can be assumed that the only entrance point of the software is by logging in.

### 3.2 Firebase Cloud Messaging service and Node.js using Firebase function

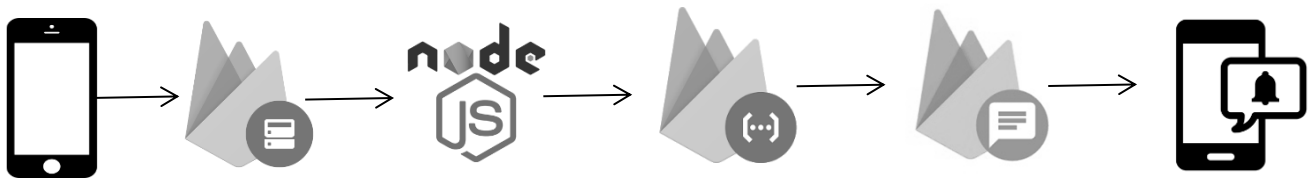

**Fig. 2.** Notification service using Firebase

In Fig. 2 the diagram says Firebase SDK helps to deliver notifications to the user with the help of Node.js where when a notification is waiting for a user in Firebase Database it trigger Node.js server to execute command and trigger command to the Firebase Function to make Firebase Cloud messaging service to send some notification to user with some data and that's how a user gets notification although he is not using the app in foreground

Normally Firebase cloud messaging service was developed to build communication between admin and users and this system is completely free for using which is a very big opportunity for us to integrate Node.js with Firebase Function to trigger Firebase cloud messaging service to send user information like chat notifications, updates as well as admin message to the users.

### 3.3 Emergency contact using Firebase SDK and Google Map SDK

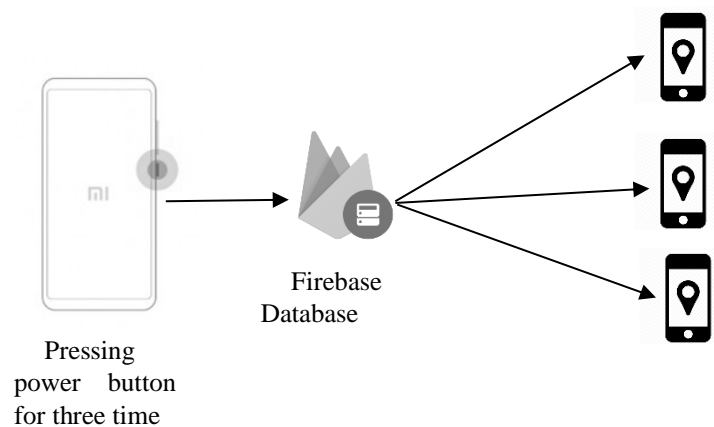

Pressing power button for three time

Firebase Database

**Fig. 3.** Emergency contact

In Fig. 3 the emergency contact feature in "SecureIT" is described with figure where a clear view to get the concept that it will send the location to the saved contacts under a user in database on power button click rapidly for 3 times.

Actually this feature is worked with the help of Google maps and Firebase SDK where firebase real-time database will store emergency contact numbers as user selected from contact or set manually using "SecureIT" and on the other hand Google Maps SDK will help the software to get the longitude and longitude of the user to send the location to the stored numbers which will take the device to the another level of security.

## 4    Google map SDK

Goggle map SDK is the main feature of our software which is truly based on Google map SDK where the software will use this SDK to get the location of the user by getting their latitude and longitude and send the location to the emergency contacts of user's choice and also show user nearby hospitals, police station and fire service on a SupportMapFragment provided by Google Map SDK itself in Android studio.

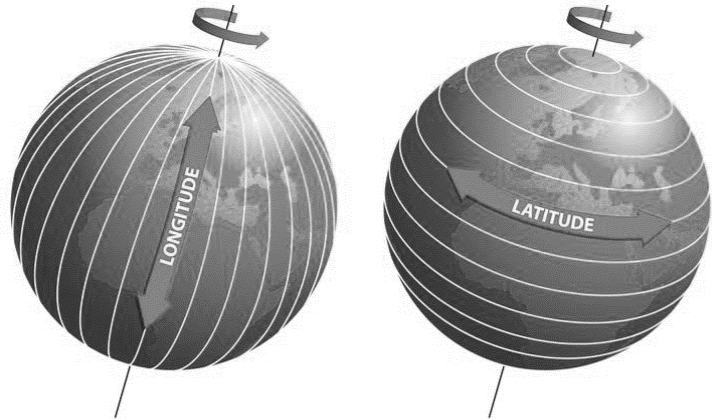

**Fig. 4.** Latitude and Longitude

Google map SDK allow user to find the current latitude and longitude of the user using Global Positioning System (GPS) and find the nearby places with the help of a well-known API for Android platform called "Google place API" which shows nearby places with custom navigation sign on SupportMapFragment provided by google after implementing Google map SDK in our project.

In Fig.  4 the way to how the software is going to see our position and nearby places by latitude and longitude with help of Google map SDK is shown. As SecureIT has the latitude and longitude of the user the software can easily pin him/her on the map which allows us both to show nearby places and also sending the location to the emergency contacts of user's choice easily.

## 5    AwesomeSplash by ViksaaSkool

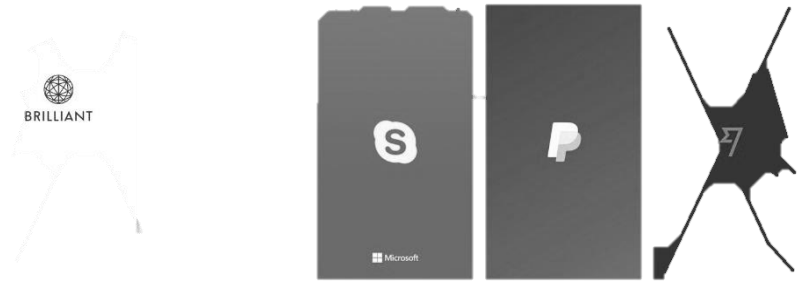

**Fig. 5.** 1  AwesomeSplash by ViksaaSkool

AwesomeSplash is an open source API developed by ViksaaSkool  available on Github which actually welcome users to the software.

AwesomeSplash is actually used for giving welcome screen of a software a material look and also make users comfortable using the software. This AwesomeSplash actually a animated welcome screen as seen in Fig.  5 which basically the first activity of the software which is also called the launcher activity.

## 6    Firebase Authentication activity

Firebase Authentication system is one of the best and secured authentication system on the web today where a user is to secured that even admin can not even know the password of the user logged in using the software. Well in "SecureIT" this very authentication is implemented in our project and now let us  see how this system works with our software.

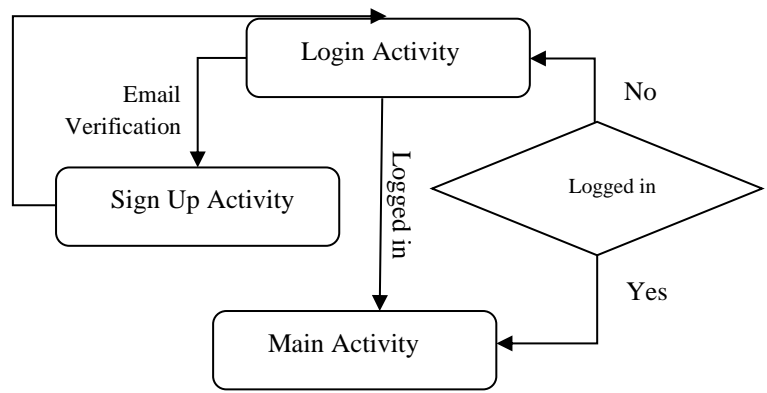

**Fig. 6.** Firebase Authentication System

In Fig. 6 the system how firebase authentication system works in this project is described.

# 7 Log in with Facebook and Google

In this project Facebook and Google API for making user feel easier to sign up during their first use were also integrated.

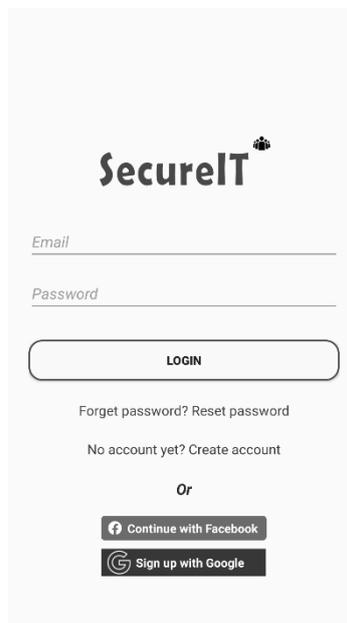

**Fig. 7.** Log in with Facebook and Google

In Fig. 7 two buttons in the activity named "Continue with Facebook" and "Sign up with Google" retrieve data like email, name and ID from Facebook or Gmail on one click and set this information in registration page and user needs only to enter the password to create an account with "SecureIT" which is developed for the comfort of the user an the usages of the software more friendly with the users of it.

## 8    Implementation

This software implantation is divided into two parts and those are Firebase    Function trigger implementation and Android based app development and those are described shortly bellow.

(1)    Firebase Function trigger implementation
-   Development tools and Languages :
Atom [10], JavaScript [11]
-   Platform :
Chrome's V8 JavaScript engine [12]
-   Libraries :
Node.js library [5]

(2)    Android app implementation
-   Development tools and Languages :
Android Studio [13], Java [14]
-   Libraries :
Google support libraries 15], Glide [16]

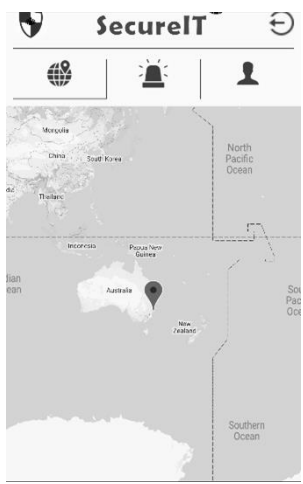

a.   Location and nearby places activity using Google map SDK and standard SupportMap-Fragment provided by Google map SDK itself [2][6]

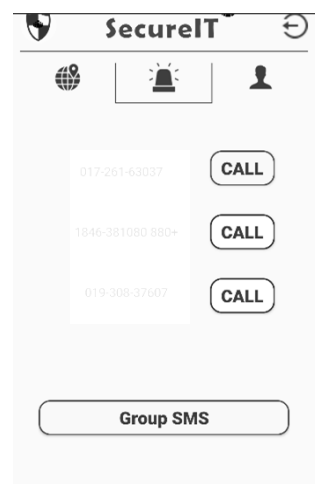

b.   Emergency contact activity using Firebase real-time database

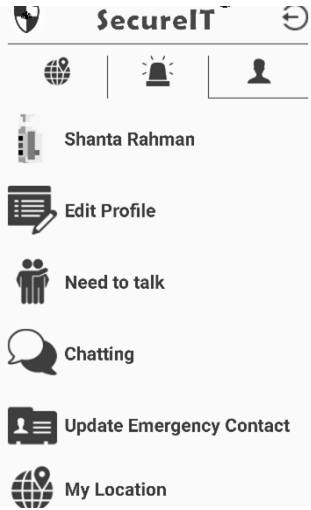

c. Profile activity

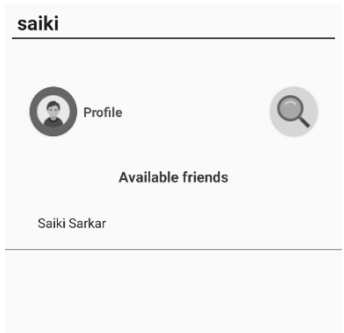

d. Search and chat activity

**Fig. 8.** Some screenshot of SecureIT

In Fig. 8 some screenshot of our project "SecureIT" is given and in (a) the activity represents a SupportMapFragment that contains google map and it will show the nearby hospitals, police stations and fire services

On the other hand (b) stands for the emergency contact which is described as Xiaomi's alternative of "Emergency SOS" and will be available to any device.

Here are some detailed functionality that are used in this software.

i. ViewFlipper : in the profile activity three different layout location(a), emergency(b) and tools(c) are included and all those layouts are controlled and functionalized by profile activity in this project. Emergency contact activity/layout is set as primary child of ViewFlippeer if user has already set the contacts in this project otherwise location is the default activity here.

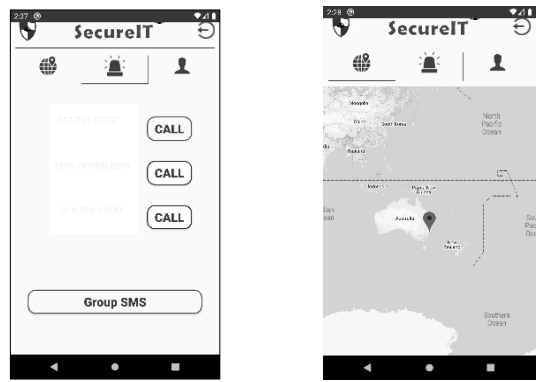

**Fig. 9.** Different state of Profile activity

In Fig. 9 left screenshot defines when a user already set the contacts in firebase real-time database and on the right side of the figure shows if the user has not set the contacts yet but still logged in using Firebase authentication system.

ii. ListView : ListView is implemented in search and chat activity(d) which shows the list of the search result based on user's search and also support click to pass shown users data like Firebase authentication UID and the name and pass it to the chat activity to continue chatting

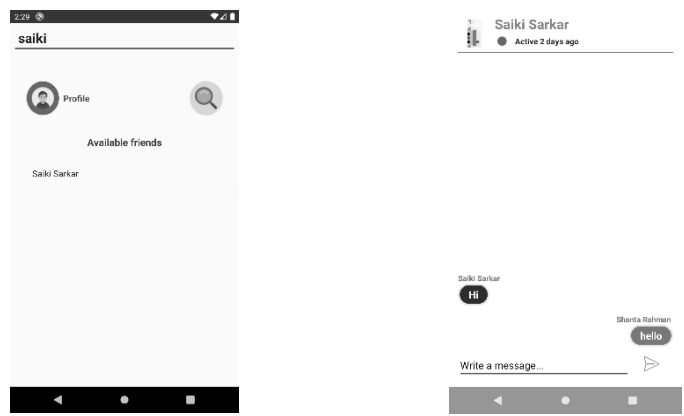

**Fig. 10.** Chatting activity

In Fig. 10 the activity shows how SecureIT provides a user friendly chatting system using firebase SDK and its usages is too simple and user friendly.

## 9    Case diagram

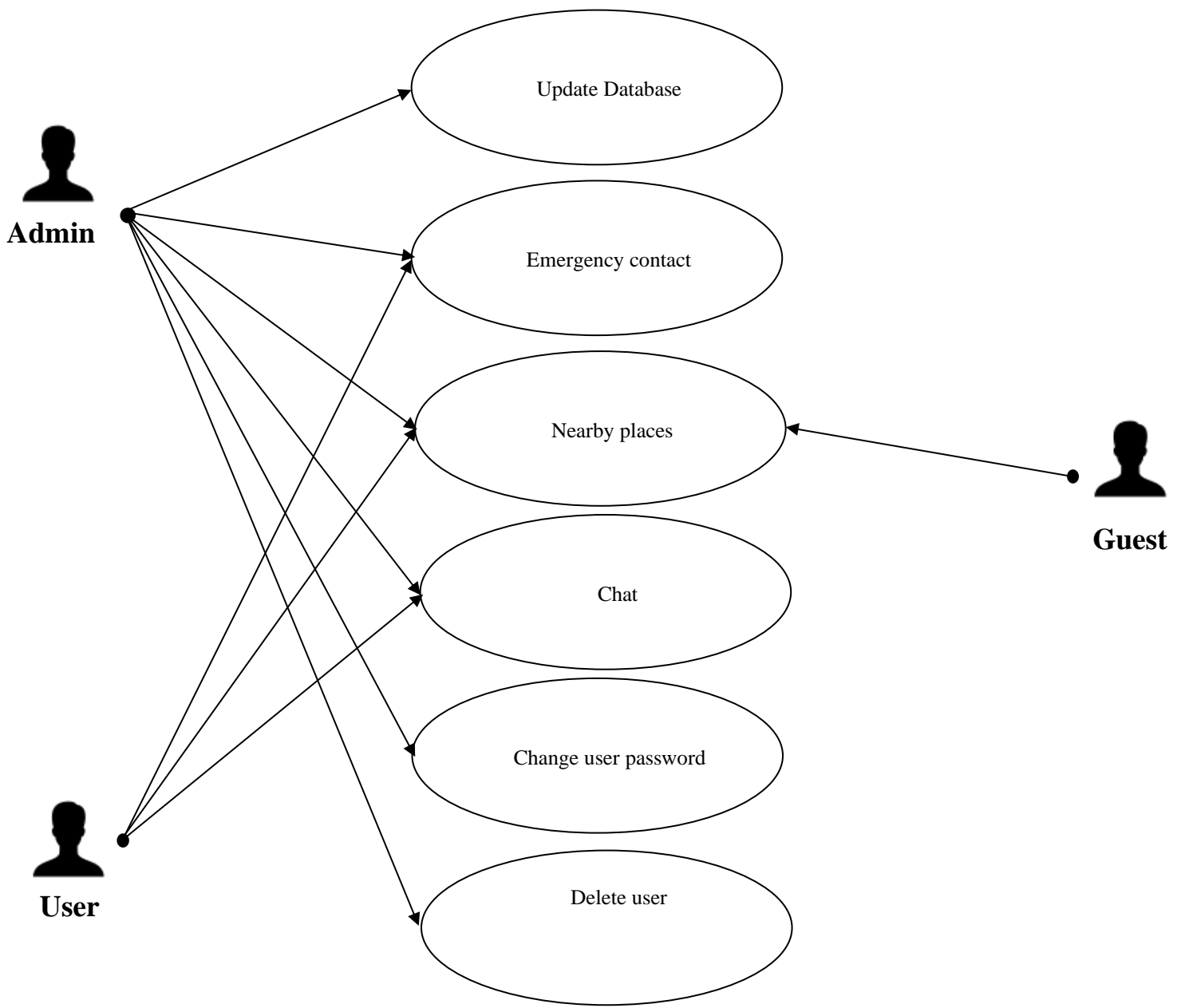

# 10    Entity relationship diagram

As we all know Firebase is not a relational database but we can use it as relational database and in this project "SecureIT" the database is used like bellow

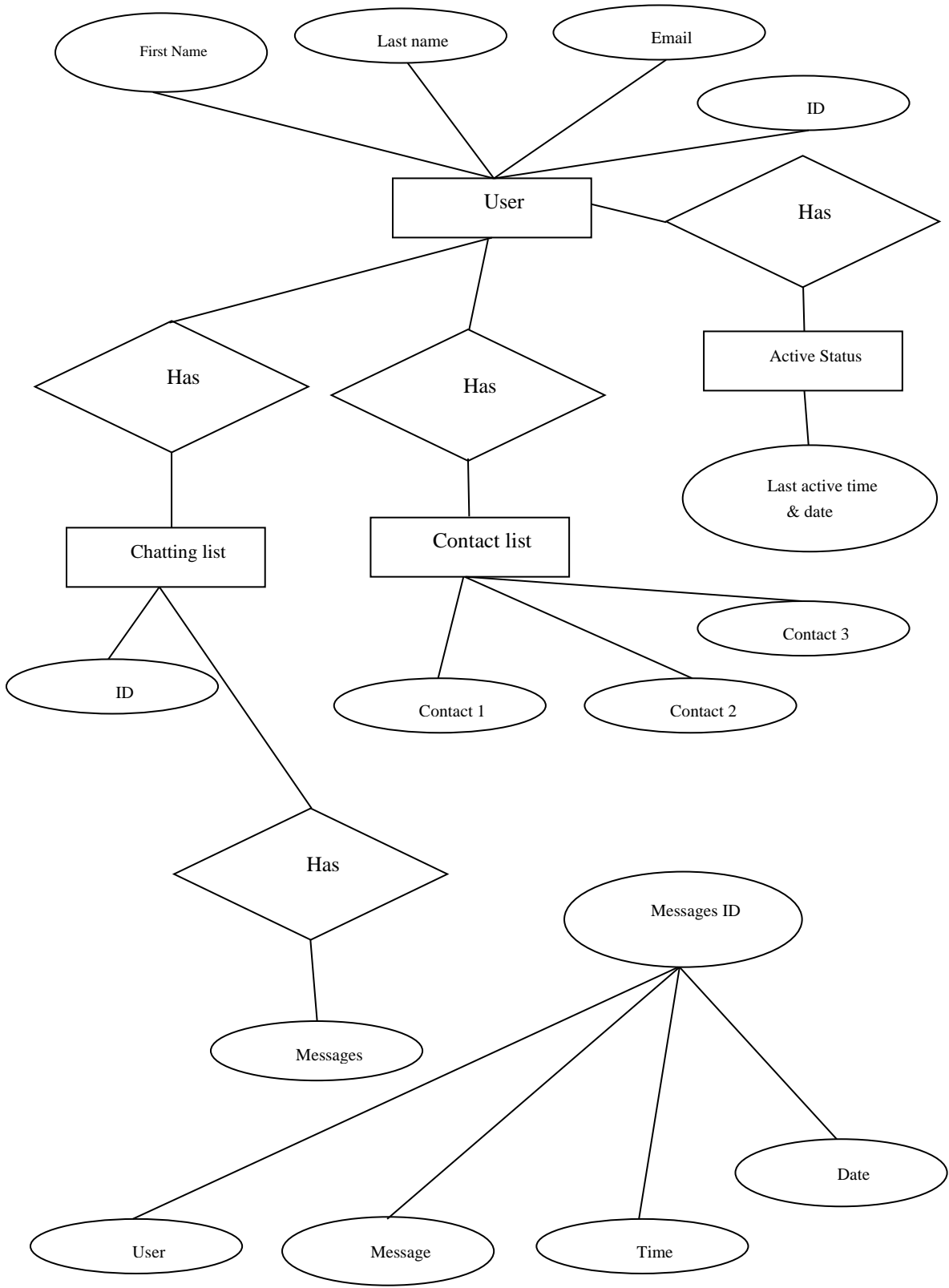

**Fig. 11.** ER diagram

In Fig. 11 we have shown the entity relationship diagram for the firebase real-time database where some detailed nodes are described below.

- User : This node contains the additional details of the user that are taken at the time of registration. It contains children like first name, last name, email and id of the user.

- Active status : In this software we set some checkpoint which will update the database about the last visit of user to that checkpoints which will help us to determine the active status of a user by getting the difference between current time and the time from the database in chat activity.

- Contact list : This node contains three contacts that are used to run emergency contact activity in our software.

- Chatting list : This node contains two child and those are the id of the people with whom user is going to chat and the messages with the individuals. Also the node messages contains message id which are pushed randomly by Firebase SDK and every message id contains messages with the time and date with the name of the user.

## Conclusion

We took a small survey on this project and also took post use review of this project from its test users and most of the review was positive. We believe this system will create a very positive impact on society and also this product will also help its users to find help in any emergency situation and also this project will send the negative side of the society back to the dark.

We know this product won't change anything in this society at where people safety was never ensured properly but we believe there is always some positive side of having something instead of having nothing.

Although the center aim of this project was to give some effort to secure women and ensure women safety but we believe this project also help every individual to be protected everywhere. Having a smartphone never declares a person to be smart and a person who is smart can gave a smartphone because they know how smartly they can use the device. So, we want people to use this software in the smartest way it can be.

## Acknowledgement

This software was planned and developed in a very short time. The developer truthfully acknowledges and thank all the reviewers as well as those people who give their helpful comments and suggestions during the development of this product.